\pdfoutput=1
\documentclass[aps,prl,twocolumn,superscriptaddress,preprintnumbers,floatfix,nofootinbib,notitlepage,showkeys,showpacs]{revtex4-1}

\usepackage[utf8]{inputenc}

\usepackage{graphicx}
\usepackage{hyperref}
\usepackage{amsmath}
\usepackage{amssymb}
\usepackage{color}

\newcommand{\be}{\begin{equation}} 
\newcommand{\ee}{\end{equation}}
\newcommand{\bea}{\begin{equation}\begin{aligned}} 
\newcommand{\eea}{\end{aligned}\end{equation}}

\newcommand{\td}{{\rm d}}

\def\lsim{\mathrel{\raise.3ex\hbox{$<$\kern-.75em\lower1ex\hbox{$\sim$}}}}
\def\gsim{\mathrel{\raise.3ex\hbox{$>$\kern-.75em\lower1ex\hbox{$\sim$}}}}

\newcommand{\Msun}{M_{\odot}}

\newcommand{\pc}{{\rm pc}}

\begin{document}

\preprint{CERN-TH-2019-112}

\title{Small-scale structure of primordial black hole dark matter \\ and its implications for accretion}

\author{Gert H\"utsi}
\email{gert.hutsi@to.ee}
\affiliation{NICPB, R\"{a}vala 10, 10143 Tallinn, Estonia}
\affiliation{Tartu Observatory, University of Tartu, Observatooriumi 1, 61602 T\~{o}ravere, Estonia}
\author{Martti Raidal}
\email{martti.raidal@cern.ch}
\affiliation{NICPB, R\"{a}vala 10, 10143 Tallinn, Estonia}
\author{Hardi Veerm\"ae}
\email{hardi.veermae@cern.ch}
\affiliation{NICPB, R\"{a}vala 10, 10143 Tallinn, Estonia}
\affiliation{Theoretical Physics Department, CERN, CH-1211 Geneva 23, Switzerland}

\begin{abstract}
Primordial black hole (PBH) dark matter (DM) nonlinear small-scale structure formation begins before the epoch of recombination due to large Poisson density fluctuations. Those small-scale effects still survive today, distinguishing physics of PBH DM structure formation from the one involving WIMP DM. We construct an analytic model for the small-scale PBH velocities that reproduces the velocity floor seen in numerical simulations, and investigate how these motions impact PBH accretion bounds at different redshifts. We find that the effect is small at the time of recombination, leaving the cosmic microwave background bounds on PBH abundance unchanged. However, already at $z=20$ the PBH internal motion significantly reduces their accretion due to the additional $1/v^6$ suppression, affecting the 21 cm bounds. Today the accretion bounds arising from dwarf galaxies or smaller PBH substructures are all reduced by the PBH velocity floor. We also investigate the feasibility for the PBH clusters to coherently accrete gas leading to a possible enhancement proportional to the cluster's occupation number but find this effect to be insignificant for PBH around $10 M_{\odot}$ or lighter. Those results should be reconsidered if the initial PBH distribution is not Poisson, for example, in the case of large initial PBH clustering.
\end{abstract}

%%%%%%%%%%%%%%%%%%%%%%%%%%%%%%%%%%%%%%%%%%%%%%%%%%%%%%%%
\maketitle

%%%%%%%%%%%%%%%%%%%%%%%%%%%%%%%%%%%%%%%%%%%%%%%%%%%%%%%%
\section{Introduction}

Primordial black holes (PBHs) can make up the entirety or a fraction of dark matter (DM) providing a viable alternative to particle DM~\cite{CHAPLINE:1975aa,Carr:2016drx,Garcia-Bellido:2017fdg}.  PBHs form from the gravitational collapse of large curvature fluctuations~\cite{Hawking:1971aa,Carr:1974nx} and can thus open a window into the very early Universe~\cite{Josan:2009qn,Cole:2017gle,Sato-Polito:2019hws}. Even when PBHs make up a small fraction of DM they may serve as seeds of galaxies~\cite{Clesse:2015wea} and supermassive black holes (BHs)~\cite{2010A&ARv..18..279V,Latif:2016qau,Bernal:2017nec} or provide an origin~\cite{Bird:2016dcv,Raidal:2017mfl,Ali-Haimoud:2017rtz,Raidal:2018bbj,Clesse:2016vqa} for the recently observed binary black hole mergers~\cite{Abbott:2016blz,Abbott:2016nmj,Abbott:2017vtc,Abbott:2017gyy,Abbott:2017oio,LIGOScientific:2018mvr}.

The abundance of PBHs is, however, constrained by several experimental observations (see, e.g.,~\cite{Carr:2016drx,Carr:2017jsz,Sasaki:2018dmp}). Recent revisions of the femtolensing~\cite{Barnacka:2012bm} and the HSC/Subaru microlensing~\cite{Niikura:2017zjd} surveys have opened the mass window $10^{-16} - 10^{-11} \Msun$ for PBH DM~\cite{Katz:2018zrn,Niikura:2019kqi}, which may be extended to even lower masses if their radiation would be modified~\cite{Raidal:2018eoo}. For higher masses PBH abundance is constrained by microlensing~\cite{Tisserand:2006zx,Allsman:2000kg,Griest:2013aaa,Zumalacarregui:2017qqd,Garcia-Bellido:2017xvr,Garcia-Bellido:2017imq,Calcino:2018mwh}, dynamics of stars in dwarf galaxies~\cite{Brandt:2016aco,Koushiappas:2017chw,Li:2016utv}, survival of wide binaries~\cite{Monroy-Rodriguez:2014ula}, gravitational wave observations~\cite{Raidal:2017mfl,Ali-Haimoud:2017rtz,Raidal:2018bbj,Authors:2019qbw,Wang:2016ana,Wang:2019kaf} and the modification of the cosmic microwave background (CMB)~\cite{Ricotti:2007au,Ali-Haimoud:2016mbv,Poulin:2017bwe} or 21 cm physics~\cite{Hektor:2018qqw,Mena:2019nhm} due to accreting PBHs.

Accretion bounds on PBHs heavier than $0.1  \Msun$  using CMB data were first obtained in Ref.~\cite{Ricotti:2007au}. These constraints were revised later in~\citep{Ali-Haimoud:2016mbv}, where it was shown that the bounds are significantly weaker constraining PBHs heavier than $100 \Msun$. These bounds depend sensitively on gas temperature, its ionization fraction and on PBH motions with respect to the gas. The motions can be broken up into several components: (i) large-scale streaming motions of the gas with respect to the DM distribution, (ii) thermal motions of the gas particles, characterized by the sound speed, and (iii) small-scale PBH motions induced by the initial discreteness noise of the PBH population. The constraints from CMB~\citep{Ali-Haimoud:2016mbv} as well as the constraints from 21 cm observations~\cite{Hektor:2018qqw,Mena:2019nhm} include components (i) and (ii) in their estimates but omit the small-scale contribution (iii). The purpose of this paper is to investigate the potential impact of component (iii)  on the allowed PBH mass fraction, $f_{\rm PBH} \equiv \Omega_{\rm PBH}/\Omega_{\rm DM}$.

After matter-radiation equality, the discreteness noise of the PBH distribution drives early small-scale structure formation, leading to the formation of binaries and an early buildup of systems with multiple PBHs. Sufficiently compact PBH systems could begin accreting coherently, which will lead to an enhanced accretion rate when compared to the case where the PBHs are treated as independent accretors. In particular, once their accretion radii start to overlap significantly, a system consisting of $N$ PBHs might start to accrete as a coherent whole, leading to an enhancement by a factor of $N$ compared to the situation with $N$ independent accretors~\citep{Lin:2007pc,Kaaz:2019wdi}. The investigation of this coherent boost factor is another task for this paper.

In this study, we make use of the small-scale $N$-body simulations in Ref.~\citep{Raidal:2018bbj}, which investigated the evolution of $30\, M_\odot$ PBHs up to redshift $z\simeq 1100$. To extrapolate the results toward lower redshifts, we build a simple analytic model for the small-scale PBH motions, which is checked against the $N$-body results at $z=1100$.

A monochromatic PBH mass function is assumed throughout the paper.

%%%%%%%%%%%%%%%%%%%%%%%%%%%%%%%%%%%%%%%%%%%%%%%%%%%%%%%%
\section{Accretion basics}

The motion of PBHs due to the Poisson enhanced small-scale structure affects mainly the constraints on PBH abundance arising from accretion of baryons during the cosmic dark ages -- the period from recombination up to the epoch of reionization -- which is the period we will focus on in this paper. Because of complexities involved with low redshift structure formation -- nonlinearities and baryonic feedback -- we will not consider the period after reionization.

\begin{figure}[t]
\begin{center}
\includegraphics[width=0.9\linewidth]{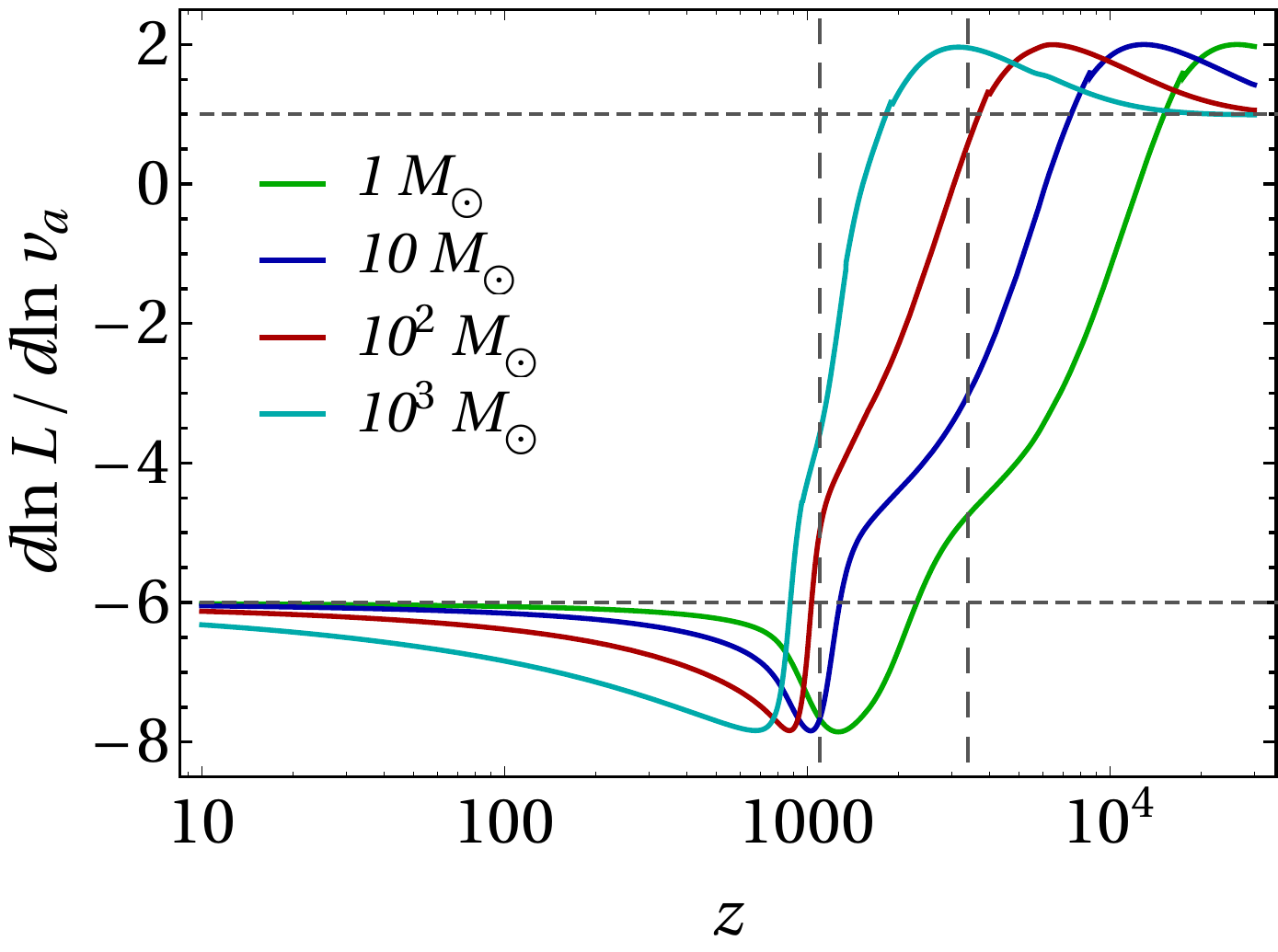}
\caption{The velocity dependence of the luminosity of PBHs for different masses at different redshifts. The luminosity was evaluated following the model in Ref.~\cite{Ali-Haimoud:2016mbv} using the substitution $v_a \to v_{\rm eff} \approx \sqrt{c_s \sigma_{\rm DM-b}}$. Nevertheless, the figure remains qualitatively the same when $v_a  = c_s$ or $v_a  = \sigma_{\rm DM-b}$ are used instead. The horizontal dashed lines indicate recombination and matter-radiation equality.}
\label{fig:dLdv}
\end{center}
\end{figure}

The accretion rate of a BH of mass $M_{\rm PBH}$ can be cast as
\be\label{eq:dotM}
	\dot M = \lambda 4\pi r_{\rm a}^2 \rho \, v_{\rm a} \propto M_{\rm PBH}^2  v_{\rm a}^{-3},
\ee
where $r_{a} \equiv GM/v_{\rm a}^2$ is the accretion radius, $v_{\rm a}$ is a characteristic velocity, and $\rho$ is the average energy density of gas. For subluminal spherical accretion, $v_{\rm a}$ is the speed of sound $c_{s}$ at infinity~\cite{Bondi:1952ni}.  The coefficient $\lambda$ is a complicated function of redshift, the characteristic velocity, and the PBH mass. For redshifts $z \lesssim 1100$ it can decrease by an order of magnitude from 1.12 to at most 0.12 with the extremes corresponding to isothermal and  adiabatic accretion, respectively~\cite{Ali-Haimoud:2016mbv}. When the motion of the BH is supersonic, then $v_{\rm a}$ is taken to be the relative velocity between the BH and the gas, i.e., $v_{\rm a} = v_{\rm rel}$, and $\lambda \approx 0.5$~\cite{10.1093/mnras/104.5.273}. An order of magnitude estimate interpolating between these two regimes can be obtained by setting
\be
	v_{\rm a} = \sqrt{v_{\rm rel}^2 + c_s^2}.
\ee
The luminosity of an accreting BH, $L = \eta \dot M$, is characterized by the radiative efficiency $\eta$, which scales roughly as $\dot M$. Thus, naively we would obtain $L \propto \dot M^2 \propto v_{\rm a}^{-6}$. However, the coefficients, e.g. $\lambda$, do also depend on $v_{\rm a}$, which can modify the velocity dependence. This dependence, obtained from the analytic model of Ref.~\cite{Ali-Haimoud:2016mbv}, is illustrated in Fig.~\ref{fig:dLdv}.  Assuming a power-law dependence at a given redshift characterized by the parameter $\kappa$, i.e., $L \propto v_{\rm a}^{\kappa}$, Fig.~\ref{fig:dLdv} shows that for the naive estimate  $\kappa = - 6$ works relatively well when $z \lesssim 1000$. If $z \gg 1000$ the scaling is qualitatively different as the velocity dependence will asymptote to $\kappa \to 1$. This transition is due to the ionization of the Universe which results in the scaling $\lambda \propto v_{\rm a}^{-3}$, characteristic of a high viscosity~\cite{Ricotti:2007jk}, as well as an extra linear dependence in the radiative efficiency. 

We remark that in Ref.~\cite{Ali-Haimoud:2016mbv} the coefficients $\lambda$  and $\eta$ were derived under the assumption of subluminal spherical accretion. However, when $z \lesssim 10^4$, accretion is mostly superluminal. By relying on the analysis in Ref.~\cite{Ali-Haimoud:2016mbv} we thus implicitly assume that the velocity dependence of $\lambda$ and $\eta$ is roughly the same in the subluminal and superluminal accretion regime.

The accretion bounds are sensitive to the injected power density, which we express as
\be\label{eq1}
	j = n_{\rm PBH}\langle L_{\rm PBH}\rangle\propto f_{\rm PBH} M_{\rm PBH}^{-\kappa/3}v_{\rm eff}^{\kappa }\,,
\ee
where $f_{\rm PBH}$ is the fraction of DM in the form of PBHs, $M_{\rm PBH}$ is the PBH mass and $v_{\rm eff}$ is an effective population-averaged velocity, 
\be\label{eq:veff}
	v_{\rm eff}^{\kappa} \equiv \langle v_{\rm a}^{\kappa} \rangle =  \int (v^2+c_s^2)^{\kappa/2}f(v)\, \td v\,.
\ee
Here $f(v)$ is the PBH velocity distribution with respect to the baryonic medium. Unless stated otherwise, we will assume $\kappa=-6$ which agrees well with the radiatively inefficient advection-dominated accretion flow model under the limit of low accretion speeds~\citep{Narayan:1994xi,Narayan:2008bv}. Since $\kappa < - 6$ during the dark ages, the choice $\kappa = - 6$ is conservative as it slightly underestimates the suppression of the luminosity effect by PBH motion. The scaling of the injected power density with the PBH mass $M_{\rm PBH}^{-\kappa/3}$ is also consistent with Ref.~\cite{Ali-Haimoud:2016mbv} as the parameters $\lambda$ and $\eta$ depend on the mass and velocity through combination $M_{\rm PBH}v_{\rm a}^{-3}$.

We will investigate how the peculiar motion of PBHs affects accretion of gas. To obtain a relative enhancement or suppression factor, it is sufficient to know the scaling relations given in \eqref{eq1}, but not the overall factor.  Thus, our conclusions will remain valid if the PBH form an accretion disk, in which case the luminosity would be enhanced but the $\kappa \approx -6$ scaling remains intact~\cite{Poulin:2017bwe}.

%%%%%%%%%%%%%%%%%%%%%%%%%%%%%%%%%%%%%%%%%%%%%%%%%%%%%%%%
\section{Motion of the PBH}

\begin{figure}[t]
\begin{center}
\includegraphics[width=\linewidth]{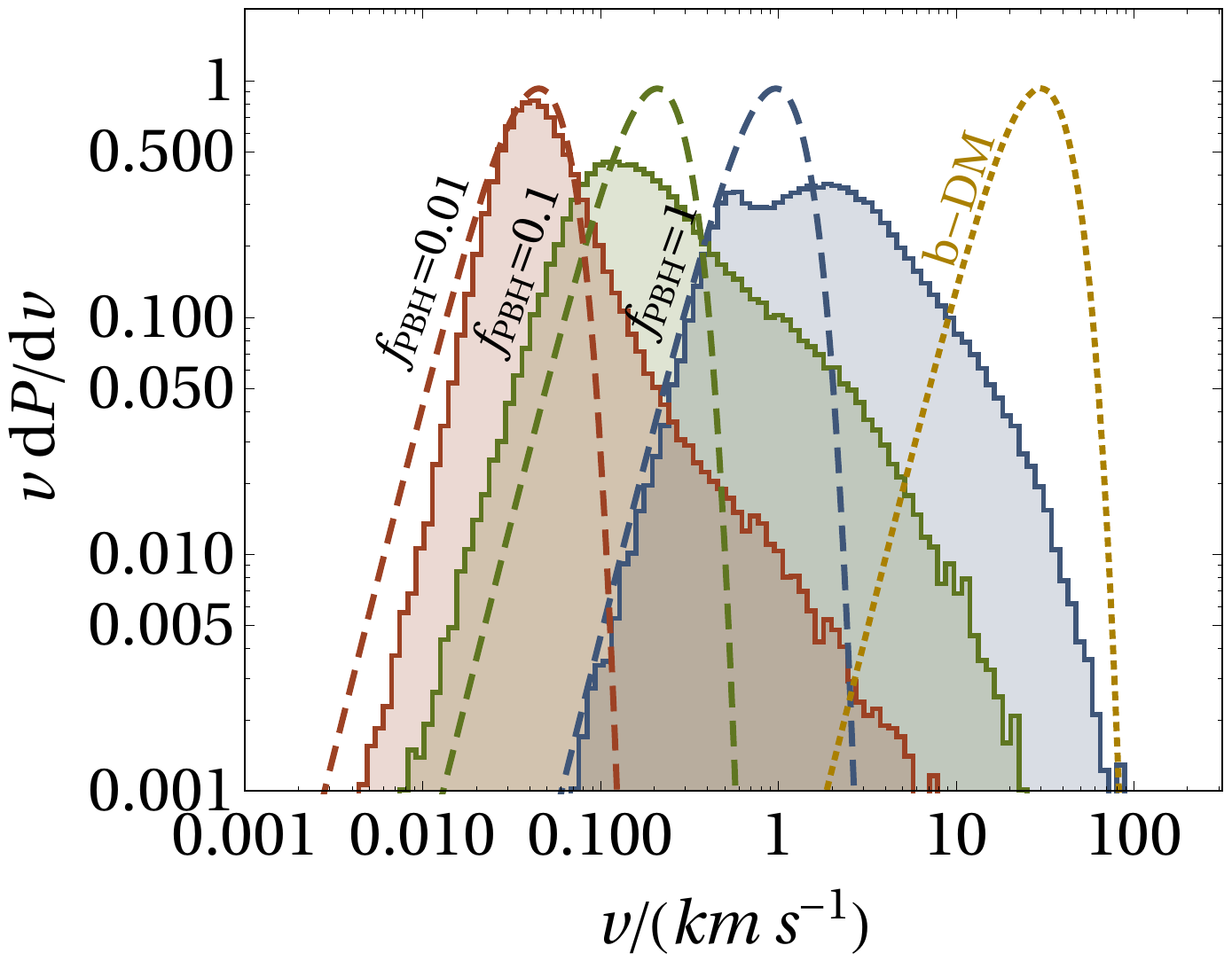}
\caption{The colored histograms show the PBH velocity distributions at redshift $z\simeq 1100$ obtained numerically in Ref.~\citep{Raidal:2018bbj}. Here $M_{\rm PBH}=30\, M_\odot$ and (from left to right) $f_{\rm PBH}=0.01,\,0.1\,{\rm and\,}1$, respectively. The dashed lines show MB distributions with dispersion \eqref{eq:sigma_v} aiming to describe the low-velocity tails of the distributions. The yellow dotted MB curve corresponds to the large-scale baryon-DM streaming motions.}
\label{fig:v_histograms}
\end{center}
\end{figure}

The velocity distribution function $f(v)$ is given by two independent components: (i) large-scale baryon-DM streaming motions described by a Maxwell-Boltzmann (MB) distribution with a 3D rms velocity $\sim 30$~km/s at recombination~\citep{Tseliakhovich:2010bj} with the usual $(1+z)$ redshift dependence, i.e., $v_{\rm rms}^{\rm DM-b}\simeq \min[1,(1+z)/10^{3}]\,30$~km/s~\cite{Dvorkin:2013cea}, and (ii) small-scale PBH motions driven by the discreteness noise of the PBH distribution. Fig.~\ref{fig:v_histograms} shows the PBH velocity distributions at redshift $z\simeq 1100$ for $M_{\rm PBH}=30\,\Msun$ PBH and for different values of $f_{\rm PBH}$ obtained from the numerical simulations of Ref.~\citep{Raidal:2018bbj}. The numerical results are compared to MB distributions (dashed lines) with the 1D velocity dispersions obtained from linear perturbation theory (see the Appendix)
\be\label{eq:sigma_v}
	\sigma_{\rm PBH}(z)\simeq 6.0\,{\rm km/s}\frac{f_{\rm PBH}^{2/3}(M_{\rm PBH}/\Msun)^{1/3}}{\sqrt{1+z}}\,.
\ee
They provide a decent fit to the low-velocity tail, which will give the dominant contribution to luminosity during the dark ages when $\kappa \approx -6$. The yellow dotted line shows the MB distribution for the baryon-DM streaming motions with 1D dispersion $\sigma_{\rm DM-b}=v_{\rm rms}^{\rm DM-b}/\sqrt{3}\simeq 17$~km/s. The low-velocity tail of the PBH velocity distribution will thus dominate over baryon-DM streaming when $1+z \lesssim 50 \,  f_{\rm PBH}^{4/9}(M_{\rm PBH}/\Msun)^{2/9}$.

The extended tail of the PBH velocity distribution can be explained by the Poisson enhanced small-scale structure, especially by the formation of PBH binaries from random close PBH pairs~\cite{Nakamura:1997sm}. The simulation in Ref.~\citep{Raidal:2018bbj} gives an overall velocity dispersion of about $f_{\rm PBH}^{2/3} \, 6.0\,{\rm km/s} $ for $30 \Msun$ PBH, which remains constant after matter-radiation equality, i.e., once the early binaries have formed. Since it is possible that $\kappa > 0$ when $M_{\rm PBH} \gtrsim 10^2 \Msun$, the presence of a high-velocity tail may \emph{enhance} the luminosity of these PBH. However, as seen in Fig.~\ref{fig:v_histograms}, the effect is present for a small fraction of PBH and, moreover, given the CMB constraint $f_{\rm PBH} \lesssim (M_{\rm PBH}/100\,\Msun)^{-2}$~\cite{Ali-Haimoud:2016mbv} and, given the velocity dispersion scales as $M_{\rm PBH}^{1/3}$, the latter will not exceed 10 km/s. In comparison to 30~km/s streaming velocities, this leads to a $< 5\%$ correction for $v_{\rm eff}$.  In conclusion, before recombination, the motion of PBHs can be safely neglected for approximate estimates even when the high-velocity tail is included.

From Eq.~\eqref{eq:sigma_v} one can see that the linearized continuity equation (see the Appendix) dictates that the PBH velocities to grow as $(1+z)^{-1/2}$, and thus, even though at the recombination epoch $\sigma_{\rm DM-b}\gg \sigma_{\rm PBH}$, at low enough redshifts $\sigma_{\rm PBH}$ is expected to dominate over $\sigma_{\rm DM-b}$, which decays as $(1+z)$.

In the following we assume that the velocity distribution function $f(v)$ in Eq.~(\ref{eq:veff}) has a MB form, i.e., $f(v)=\sqrt{2/\pi}(v^2/\sigma^3)\exp(-v^2/(2\sigma^2))$, with a 1D dispersion $\sigma=(\sigma_{\rm DM-b}^2+\sigma_{\rm PBH}^2)^{1/2}$. If $\sigma_{\rm DM-b}\gg \sigma_{\rm PBH}$ the MB assumption is completely fine. In the case $\sigma_{\rm PBH}\gg \sigma_{\rm DM-b}$ it is also a good approximation, since the integral in Eq.~\eqref{eq:veff} is determined by the low-velocity tail, which is well approximated by the MB distribution. For $\sigma_{\rm PBH}\sim \sigma_{\rm DM-b}$, this approximation somewhat underestimates the effective $\sigma$. Our results for the accretion bounds are therefore conservative. A more realistic and precise treatment demands a full model for the PBH velocity distribution along with its temporal evolution. 

For the sound speed $c_s$ we use the approximation
\bea\label{eq:cs}
	c_s^2(a)	&=\frac{\gamma k_{\rm B}}{\mu m_p}T_k(a)\,,\\
	T_k(a)	&=\frac{T_{\rm CMB}}{a}\left[1+\frac{a/a_1}{1+(a_2/a)^{3/2}}\right]^{-1}\,,
\eea
where $T_k$ is the gas kinetic temperature, $a$ is the scale factor, $\gamma=5/3$ is the adiabatic index and $\mu\simeq 1.22$ is the mean atomic weight of the neutral gas in units of the proton mass $m_p$.
$T_k(a)$ has the analytic fitting form suggested in~\citep{Tseliakhovich:2010bj}. Here $T_{\rm CMB}=2.725$~K is the CMB temperature at $z=0$~\citep{Fixsen:2009ug} and $a_1=1/136$ and $a_2=1/181$.\footnote{Our values for $a_1$ and $a_2$ differ from the values given in~\citep{Tseliakhovich:2010bj} because we use somewhat different $\Lambda$CDM parameters, $\Omega_m=0.3$, $\Omega_b=0.05$ and $h=0.7$.} The above approximation for $T_k$ is accurate $\lesssim 4\%$ when compared against the numerical results from the RECFAST code~\citep{Seager:1999bc}.
During the dark ages $\sigma_{\rm DM-b} > 2 c_s$, that is, the streaming velocity dominates for most PBH. When $c_s \ll \sigma$, then $v_{\rm eff} \approx 1.03 \sqrt{c_s \sigma}$.

%%%%%%%%%%%%%%%%%%%%%%%%%%%%%%%%%%%%%%%%%%%%%%%%%%%%%%%%
\section{Coherent accretion boost}

\begin{figure}[t]

\begin{center}
\includegraphics[width=0.9\linewidth]{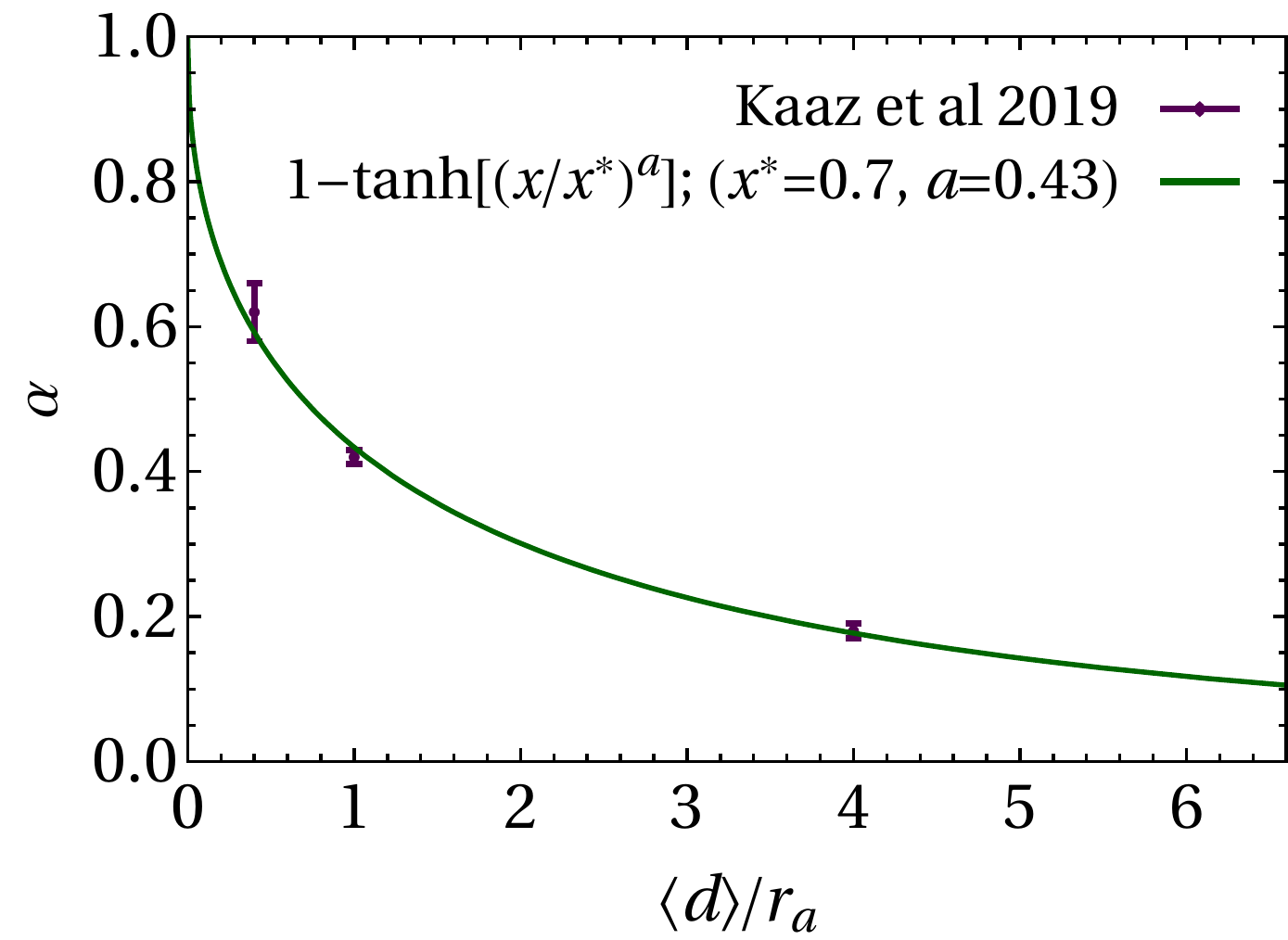}
\caption{Power-law index $\alpha$ for the coherent accretion boost as a function of the ratio between the mean intracluster PBH distance $\langle d \rangle$ and an effective accretion radius $r_{\rm a}$. The points with error bars show the results from numerical simulations~\citep{Kaaz:2019wdi}. The green solid line represents our analytic fit used throughout this paper.}
\label{fig:alpha}
\end{center}
\end{figure}

The total accretion rate of a cluster comprising $N$ PBHs scales as $\dot{M}_{\rm tot}\propto N$ when all cluster members accrete independently. This approximation is valid when the average distance between the PBHs is significantly larger than their effective accretion radius. At the other extreme, that is, once the accretion radii start to overlap significantly, the cluster begins to accrete as a coherent whole. In this case $\dot{M}_{\rm tot}\propto N^2$, that is, the accretion is enhanced by a factor of $N$~\citep{Lin:2007pc,Kaaz:2019wdi}. Between these extremal cases, the enhancement factor over the standard incoherent feeding can be approximated as $N^\alpha$, where $0 < \alpha < 1$. The exponent $\alpha$ depends on the ratio of mean PBH distance inside the clusters $\langle l \rangle$ to the accretion radius $r_{\rm a}$. We use the functional form for $\alpha$ shown in Fig.~\ref{fig:alpha}, which is motivated by the simulation results of Ref.~\citep{Kaaz:2019wdi}.~\footnote{Note the difference by a factor of 2 in our definition of $r_a$ when compared to Eq.~(4) of Ref.~\citep{Kaaz:2019wdi}.}

To obtain the mean PBH distance inside clusters we make a standard assumption that clusters are identified as objects with overdensities $\Delta$ times over the background matter density. For the redshifts considered in this paper $\Delta$ is well approximated by its standard Einstein-de Sitter value, $\Delta=18\pi^2$.
Under these assumptions, the average proper PBH distance inside clusters is
\bea\label{eq:d_av}
\langle d \rangle
&	=\frac{1}{1+z}\left[\frac{3M_{\rm PBH}}{4\pi \times 18\pi^2 f_{\rm PBH} \Omega_{\rm m} \rho_c}\right]^{1/3} \\
&	\simeq \frac{32\,\pc}{1+z}\left[\frac{M_{\rm PBH}}{f_{\rm PBH} \Msun}\right]^{1/3}\,.
\eea

The accretion radius will vary for different clusters depending on the relative motion between the PBH and the gas. To estimate the effect of motion consider the luminosity-weighted average, with weight $w \propto L \propto v_{\rm a}^{\kappa}$
\be
	\langle r_{\rm a} \rangle_{L}
	\equiv GM_{\rm PBH} \frac{\langle v_{\rm a}^{\kappa-2} \rangle}{\langle v_{\rm a}^{\kappa} \rangle}
	\approx \frac{3+\kappa}{\kappa} \frac{GM_{\rm PBH}}{c_s^2}
\ee
which for $\kappa = -6$ is half the accretion radius of a PBH at rest. The approximation is valid when $c_s \ll \sigma$ and $\kappa < -3$ and can thus be used during the dark ages. The dependence on the PBH velocity distribution drops out since the luminosity is dominated by the slowest PBHs. 

As the accretion radius decreases faster with redshift than the average distance, the largest coherent boost is expected at smaller redshifts. The speed of sound \eqref{eq:cs} is approximated by  $c_s \approx (1+z)  \times 15 \rm \, m/s $ around $z=10$. The condition $\langle r_{\rm a} \rangle_{L} \gtrsim \langle d \rangle$ is satisfied when
\be
	m_{\rm PBH} \gtrsim 200 \Msun \left(\frac{1+z}{10}\right)^{3/2} f_{\rm PBH}^{-1/2}.
\ee
So, as the abundance of the PBH with masses over $100 \Msun$  are strongly constrained, coherent accretion of PBH is suppressed in viable mass ranges. An exception is possible in the central regions of halos where the density can be much above the average or for the small clusters where the average distance can drop by an order of magnitude (see Fig.~\ref{fig:Mfrac}). For binaries, the large peculiar velocity will suppress accretion when compared to the individual BHs. This effect is, however, milder for highly eccentric binaries~\cite{Postnov:2017nfw}. 

The effect of coherent accretion on injected energy is shown in Fig.~\ref{fig:correction}. At $z=10$ it can be sizable already for $100 \Msun$ PBH. This is partly due to the large $N$ of PBH clusters at $z=10$. {In the allowed mass range $m_{\rm PBH} \leq 100 \Msun$, the effect tends to be small. It can be further suppressed due to the evaporation of small PBH clusters~\cite{Afshordi:2003zb} and due to the internal virial motion of the PBH in the clusters that can reduce the accretion radius especially in the outer region of the cluster, effectively reducing its size. However, a more accurate understanding of accretion into such clusters likely requires a numerical approach that accounts for the density profile as well as for the internal motion of PBHs in the cluster.}

%%%%%%%%%%%%%%%%%%%%%%%%%%%%%%%%%%%%%%%%%%%%%%%%%%%%%%%%
\section{Results}

\begin{figure}[t]
\begin{center}
\includegraphics[width=0.95\linewidth]{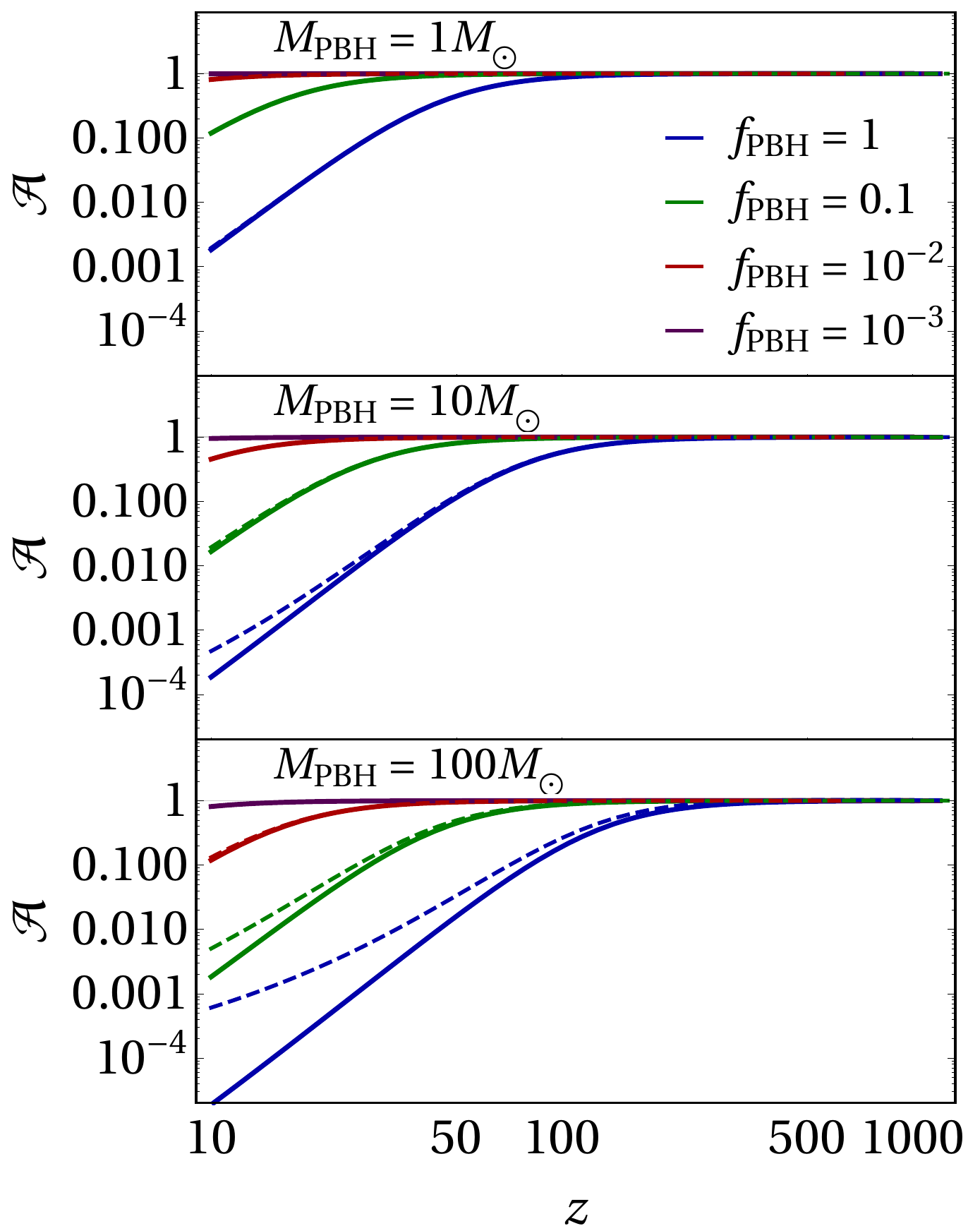}
\caption{The solid lines show the energy input modification factors \eqref{factor} for a range of redshifts without the coherent accretion enhancement neglected ($\alpha = 0$). The dashed lines also account for the coherent accretion boost. The suppression toward low redshifts is driven by the PBH motions, whereas possible enhancement due to coherent accretion effect is sizable only for $\gtrsim 100\,\Msun$ PBH and $f_{\rm PBH} \sim 1$. Top, middle and bottom parts correspond to $M_{\rm PBH}=1,\,10\,\,{\rm and}\,100\,\Msun$, respectively, while $f_{\rm PBH} \in \{0.001, 0.01,0.1,1\}$.}
\label{fig:correction}
\end{center}
\end{figure}

The modification of the energy input by accretion is given by the factor
\be\label{factor}
	\mathcal{A}  
	\equiv \frac{\langle L \rangle}{\langle L \rangle_{\rm std}} 
	= \frac{\langle N^{\alpha} v_{\rm a}^{-6}\rangle}{\langle v_{\rm a}^{-6}\rangle_{\rm std}}
\ee
where the superscript "$\rm std$" represents the ``standard'' case with a uniform PBH distribution and without PBH peculiar motions. The average is also taken over the cluster mass distribution for which we use a discretized Press-Schechter-like halo mass function (see the Appendix, Eq.~\eqref{eq:Nav}). This is justified because the coherent accretion boost is active during the late dark ages where the cluster size can be relatively large and when the smallest PBH clusters have been evaporated due to their relatively short dynamical timescales.  The results for the factor $\mathcal{A} $ are presented in Fig.~\ref{fig:correction} for different values of $f_{\rm PBH}$ and $M_{\rm PBH}$, with and without coherent accretion boost factor $N^\alpha$. When the latter is omitted, then \eqref{factor} is approximated by $\mathcal{A} \approx (1+ \left(\sigma_{\rm PBH}/\sigma_{\rm DM-b}\right)^2)^{-3/2}$. The effect of the $N^\alpha$-term can be neglected for order of magnitude estimates when $f_{\rm PBH} \lesssim 0.1$ or $M_{\rm PBH} \lesssim 10$ and the dominant contribution is driven by the $v_{\rm eff}$ dependence.

\begin{figure}[t]
\begin{center}
\includegraphics[width=0.95\linewidth]{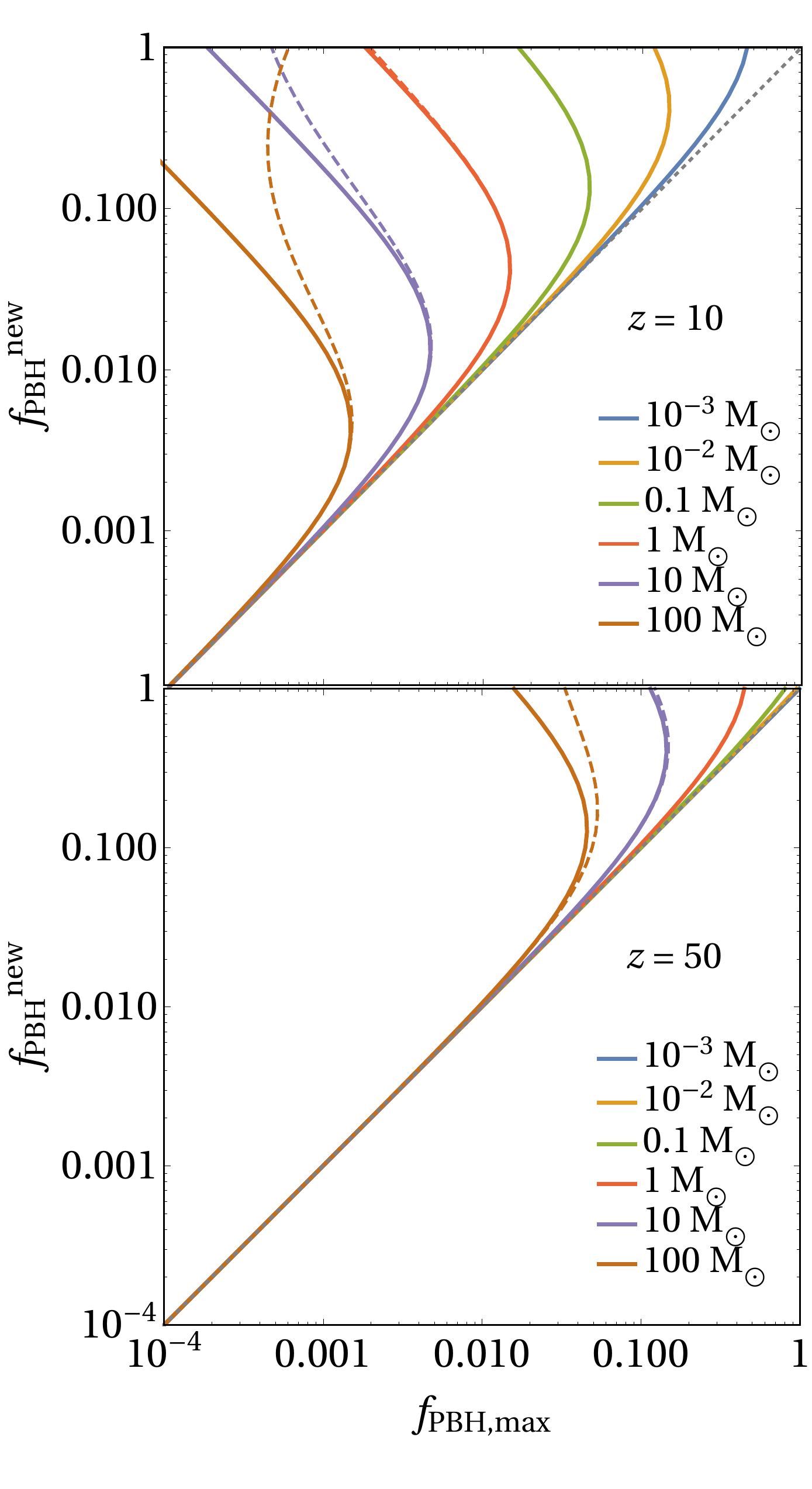}
\caption{$f_{\rm PBH}^{\rm new}$-$f_{\rm PBH}$ relation for various values of $M_{\rm PBH}$. Top/bottom panel corresponds to redshift $10/50$. The dashed lines include the effect of coherent accretion while the solid lines ignore it.}
\label{fig:f_pbh}
\end{center}
\end{figure}

Consider now the effect on constraints on PBH abundance. If the accretion bound without discreteness-induced effects is $f_{\rm PBH, max}$, the corrected bound can be obtained by comparing the injected energy densities. This gives
\be
	f_{\rm PBH, max} = \mathcal{A}(f_{\rm PBH}^{\rm new}) f_{\rm PBH}^{\rm new},
\ee
assuming the injected power density in \eqref{eq1} scales linearly with $f_{\rm PBH}$ when the PBH motion is omitted. The modified boundary of the allowed region $f_{\rm PBH}^{\rm new}$ is plotted in Fig.~\ref{fig:f_pbh} for redshifts $10$ and $50$ for various values of $M_{\rm PBH}$. The larger the value of $M_{\rm PBH}$ the more $f_{\rm PBH}^{\rm new}$ deviates from the bound without PBH motions. It is interesting to note the nonmonotonic behavior of the $f_{\rm PBH}^{\rm new}$-$f_{\rm PBH}$ relation. This is easily understood with Eq.~\eqref{eq1}, according to which an increase of $f_{\rm PBH}$, when it is small, leads to an increase in the injected power density $j$. However, this increase in $j$ will saturate once $f_{\rm PBH}$ becomes sufficiently large for the PBH motions to become noticeable.

The modification on the PBH abundance bounds is shown is Fig.~\ref{fig:f_pbh}. For example, according to Fig.~\ref{fig:f_pbh}, for $M_{\rm PBH}=100\,M_\odot$ and $z=10$, if, neglecting PBH motions, the bound on $f_{\rm PBH}$ is $f_{\rm PBH}\lesssim 0.001$, then, after accounting for the corrections, the unconstrained region consists of two  regions $f_{\rm PBH} \lesssim 0.001$ and $f_{\rm PBH} \gtrsim 0.01$, i.e., a new allowed region emerges. For the slightly weaker noncorrected constraint, e.g., $f_{\rm PBH}\lesssim 0.002$, however, all the values of $f_{\rm PBH}^{\rm new}$ are allowed.

%%%%%%%%%%%%%%%%%%%%%%%%%%%%%%%%%%%%%%%%%%%%%%%%%%%%%%%%
\section{Discussion and summary}

We have investigated (i) the effect of PBH motions on energy injection from the gas accretion along with (ii) possible boost due to coherent accretion inside PBH clusters. We find that the first effect has a dominant impact, while the second one can be neglected for PBH masses and abundances allowed by the constraints.

At redshifts $z\sim 1000$ the PBH motions are still small compared to the dominant baryon-DM streaming motions and thus the impact on existing CMB bounds (e.g.~\citep{Ali-Haimoud:2016mbv}) is negligible. However, at redshifts $z\lesssim 100$ the effect of PBH motions should certainly be included. Without it one ends up with unrealistically (up to several orders of magnitude) tight bounds as can be seen in Figs.~\ref{fig:correction} and~\ref{fig:f_pbh}.

We should point out that in this paper the discreteness-induced PBH motions were estimated via a linearized continuity equation, which was shown to capture the early evolution quite well. In particular, the low-velocity tail of the PBH velocity distribution, where most of the energy input from accretion is released, turned out to be well described through this simplified treatment. It is clear that nonlinear evolution adds extra small-scale motions, resulting in an even stronger impact on deducible accretion bounds. Thus, our treatment here is certainly on the conservative side. A complete treatment here calls for a dedicated numerical simulation, which is beyond the scope of this paper.

We assumed a monochromatic PBH mass distribution and that the primordial spatial distribution is Poisson. In case the PBHs are clustered, i.e., there are nontrivial primordial correlations between PBHs, both the coherent enhancement and the correction due to PBH motion are expected to increase, as the formation of structure begins earlier. In this case even the CMB bounds may be tightened, especially for heavier PBH for which the luminosity can grow when the velocity is increased. An extended mass function will nontrivially affect the evolution of PBH clusters because the heavier PBH tend to migrate toward the center of such clusters and are more likely to form hard binaries while disrupting the lighter ones. In particular, once the small-scale effects become relevant, one cannot use the method of, e.g.,~\cite{Carr:2017jsz} to obtain constraints for extended mass functions.

The main message of this work is that discreteness-induced PBH motions must be accounted for in case one wishes to derive reliable accretion bounds in the late Universe. Thus, e.g., the 21 cm PBH bounds derived in~\cite{Hektor:2018qqw,Mena:2019nhm}, which neglect the above motions, need to be appropriately adjusted. In the wider context, these discreteness-driven motions provide an inescapable lower velocity floor which throughout can only grow as it evolves. Therefore, e.g., compared to the standard CDM case the halos cannot have very cool central regions; also, the early stages of structure formation ($z\sim {\rm few}\times 10$), during which the shot-noise-driven PBH motions have considerable impact, differ significantly. To investigate these issues in greater detail, one must rely on dedicated numerical simulations. These will certainly be quite demanding, ideally requiring mass resolutions many orders of magnitude below the resolutions of typical cosmological simulations, since only then can one properly account for the physical PBH shot-noise level. A corresponding CDM simulation against which to compare the PBH run needs an even higher resolution.

%%%%%%%%%%%%%%%%%%%%%%%%%%%%%%%%%%%%%%%%%%%%%%%%%%%%%%%%
\section*{Acknowledgements}
This work was supported by the grants IUT23-6, IUT26-2, by EU through the ERDF CoE program grant TK133,  and by the Estonian Research Council via the Mobilitas Plus grant MOBTT5. 

%%%%%%%%%%%%%%%%%%%%%%%%%%%%%%%%%%%%%%%%%%%%%%%%%%%%%%%%
\appendix

\section{Appendix: structure formation estimates}

In this section, we provide simple estimates for the halo mass function resulting from the initial uniform spatial Poisson distribution using the Press-Schechter (PS) formalism~\cite{Press:1973iz}. PS estimates work for sufficiently large scales, such that the initial granularity of the PBH density field can be neglected. To extrapolate the formalism to smaller scales, we treat the initial discrete Poisson shot-noise field as an equivalent continuous white noise field. We also estimate the level of large-scale bulk motions induced by the PBH density fluctuation field.

In the early Universe small-scale fluctuations in the matter density field are dominated by the PBH discreteness fluctuations with a flat power spectrum
\be
	P_{\rm PBH}(z)=g^2(z)\left(\frac{\Omega_{\rm PBH}}{\Omega_m}\right)^2\frac{1}{n_{\rm PBH}}\,,
\ee
where $g(z)$ is the linear growth factor, $n_{\rm PBH}$ is the comoving PBH number density, and $\Omega_{\rm PBH}$ and $\Omega_m$ are the PBH and matter density parameters, respectively. Before the matter-radiation equality, the fluctuation growth is only logarithmic, so one can neglect it. In the following, we use only the growth factor relevant for the matter-dominated Universe $g(z)\propto (1+z)^{-1}$. Under these assumptions the PBH power spectrum can be expressed as
\be\label{pbh_p}
P_{\rm PBH}(z)\simeq \left(\frac{1+z_{\rm eq}}{1+z}\right)^2\frac{1-f_b}{\Omega_m\rho_c}f_{\rm PBH}M_{\rm PBH}\,.
\ee
Here the equality redshift $z_{\rm eq}\simeq 3400$, the baryon fraction $f_b\equiv \Omega_b/\Omega_m\simeq 1/6$, the DM fraction in the form of PBHs $f_{\rm PBH}\equiv \Omega_{\rm PBH}/\Omega_{\rm dm}$, $M_{\rm PBH}$ is the PBH mass and $\rho_c$ is the critical density.

\subsubsection{Halo mass function}

Within the PS formalism, the PBH fluctuation field with the above-given flat spectrum will lead to the following halo mass function:
\bea\label{hmf}
	\frac{{\rm d}n}{{\rm d}\ln M}(M,z)
	&=		\frac{n_{\rm PBH}}{\sqrt{\pi}}\frac{M_{\rm PBH}}{M}\left[\frac{M}{M_*(z)}\right]^{1/2}\nonumber\\
	&\times	\exp\left[-\frac{M}{M_*(z)}\right]\,,
\eea
where the characteristic halo mass
\bea\label{eq_mstar}
	M_*(z)
	&\equiv	\frac{2}{\delta_c^2}\left(\frac{1+z_{\rm eq}}{1+z}\right)^2(1-f_b)f_{\rm PBH}^2M_{\rm PBH}\nonumber\\
	&\simeq	\left(\frac{2600}{1+z}\right)^2f_{\rm PBH}^2M_{\rm PBH} \equiv N_*(z) M_{\rm PBH}\,.
\eea
Here $\delta_c\simeq 1.686$ is the critical density for spherical collapse. At the recombination epoch, $z\sim 1100$, this gives $M_*\sim 6 M_{\rm PBH} f_{\rm PBH}^2$; thus for $f_{\rm PBH} \leq 0.6$, we expect that the PBH will not form structures beyond binaries.

\begin{figure}[t]
\begin{center}
\includegraphics[width=0.95\linewidth]{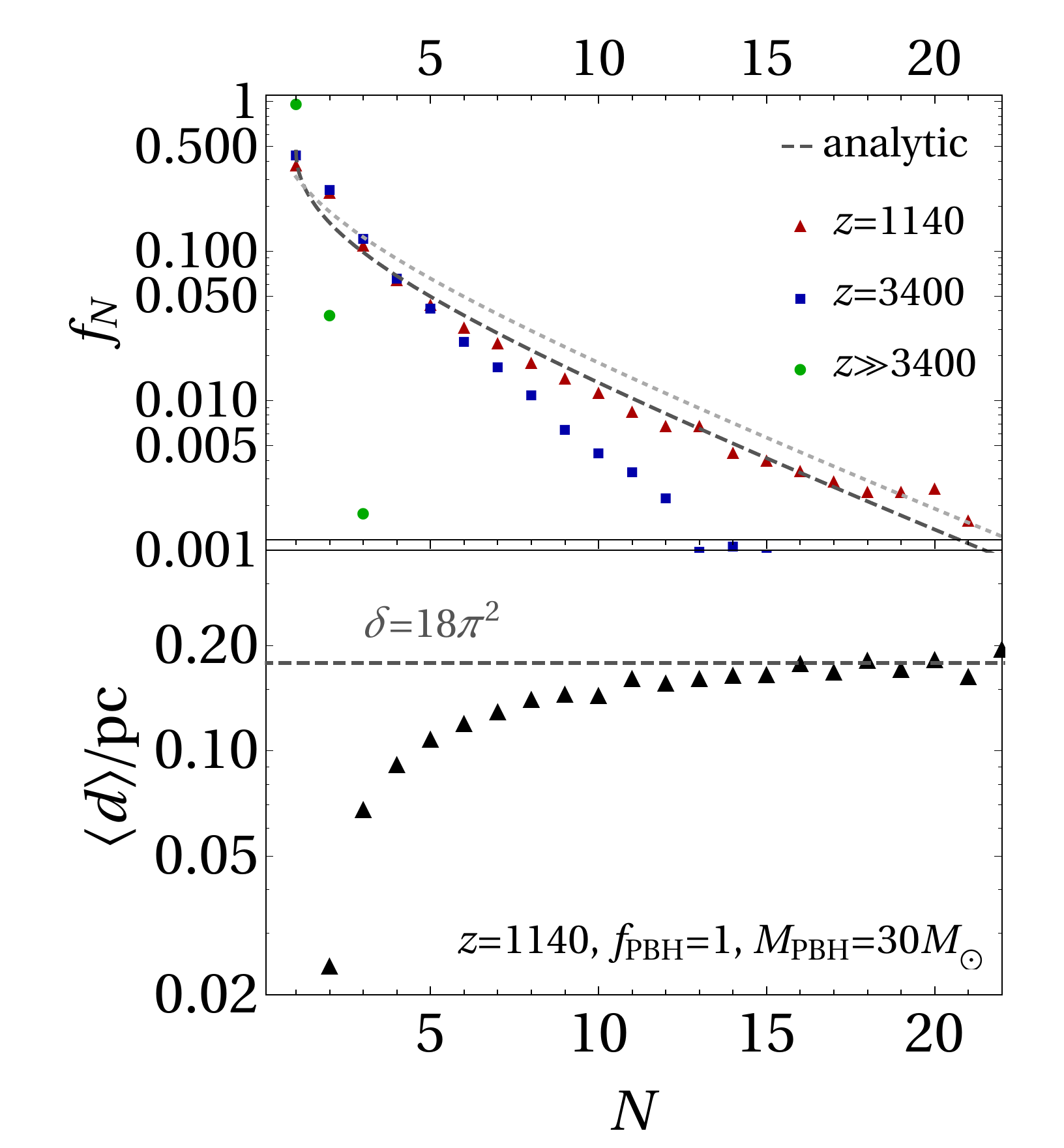}
\caption{{\it Upper panel: }Fraction of mass in halos as a function of PBH occupation number, assuming $f_{\rm PBH}=1$ and $z=1140$. Data points show the results from $N$-body simulations of Ref~\citep{Raidal:2018bbj} with clusters found by a friends-of-friends algorithm. The solid line is given by the analytic estimate Eq.~\eqref{eq_mfrac}, which due to $M_*\propto M_{\rm PBH}$ does not depend on $M_{\rm PBH}$. {\it Lower panel:} Average distance defined as $R_{c} N^{-1/3}$, where $R_{c}$ is the size of the cluster, averaged over clusters with a given number of PBH, $N$. The darker dashed line shows the expectation from the discretized PS formalism \eqref{eq_mfrac} and the light dotted line corresponds to Eq.~\eqref{eq_mfrac2} normalized to unity. }
\label{fig:Mfrac}
\end{center}
\end{figure}

In the above calculations the halo mass was a continuous parameter. Here we use a simple method to convert to a corresponding discrete case, where the halo mass can take values $M_N=NM_{\rm PBH},\ N\in\mathbb{N}$. In particular, we are interested in calculating a total mass fraction in halos with mass $M_N$. Integrating Eq.~(\ref{hmf}) over halo mass $M$ gives us a total mass density in PBHs, $\rho_{\rm PBH}\equiv n_{\rm PBH}M_{\rm PBH}$, and so the corresponding mass probability distribution function is given by Eq.~(\ref{hmf}) divided with $\rho_{\rm PBH}$. To obtain a discrete version of this probability distribution, $f_N$,we approximate a contribution from halos with mass $M_N$ by integrating from $(N-1)M_{\rm PBH}$ to $NM_{\rm PBH}$. This gives
\be\label{eq_mfrac}
	f_N(z) = {\rm erf}\left(\sqrt{\frac{N}{N_*(z)}}\right)-{\rm erf}\left(\sqrt{\frac{(N-1)}{N_*(z)}}\right)\,.
\ee
By construction, this probability distribution is correctly normalized, i.e., $\sum_{N=1}^{\infty}f_N=1$, and it reduces to the usual PS formalism for large $N$,
\be\label{eq_mfrac2}
	f_N(z) \propto \sqrt{\frac{N_*(z)}{N}} \exp\left(-\frac{N}{N_*(z)}\right)\,.
\ee
The comparison of this analytic approximation against the $N$-body results of~\citep{Raidal:2018bbj} are shown in Fig.~\ref{fig:Mfrac} for $f_{\rm PBH}=1$ and $z= 1100$. Although the analytic cluster mass function shows a good agreement with the numerical data, we see in Fig.~\ref{fig:Mfrac} that the average PBH separation in small clusters can be much larger than expected from the PS formalism. 

When estimating the coherent accretion boost we will use \eqref{eq_mfrac2} as the approximation that gives
\be\label{eq:Nav}
	\langle N^{\alpha} \rangle \approx \frac{ {\rm Li}_{1/2 - \alpha}\left[ \exp\left(-1/N_*(z)\right) \right] } { {\rm Li}_{1/2}\left[ \exp\left(-N_*(z)\right) \right] }.
\ee
This expression works well when PBHs form larger clusters, i.e. when $N_*(z) \gg 1$, which is satisfied for viable $f_{\rm PBH}$ and $M_{\rm PBH}$ when $z \ll 1000$.

\subsubsection{Linear peculiar velocity field}
To estimate the evolution of the low-velocity tail of the velocity distribution, as characterized by~\eqref{eq:sigma_v}, we start from the linear continuity equation
\be
	\dot{\delta_k}+ikv=0\,,
\ee
which can be rewritten as
\be
	v=-iaf(a)H(a)g(a)\frac{\delta_k(a=1)}{k}\,,
\ee
and the peculiar velocity power spectrum can be recast as
\be
	P_v(k,a)=\left[af(a)H(a)g(a)\right]^2\frac{P(k,a=1)}{k^2}\,.
\ee
Above, $v$ is the peculiar velocity component parallel to the wave vector ${\bf k}$, $\delta_k$ is the Fourier component of the density fluctuation, $a$ is the scale factor (normalized such that $a=1$ at $z=0$), $H(a)\equiv \dot{a}/a$ is the Hubble parameter, $g(a)$ is the linear growth factor, and $f(a)$ is the dimensionless linear growth rate $f(a)\equiv {\rm d}\ln g(a)/{\rm d}\ln a$.

The dispersion of the velocity fluctuation field smoothed over the comoving scale $R$ with filter $W$ can be expressed as
\be\label{eq:sigma_v_def_app}
	\sigma_v^2(R,z) = \frac{1}{2\pi^2}\frac{f^2(z)H^2(z)}{(1+z)^2}\int P(k,z)W^2(kR)\, \td k \,.
\ee
The spectrum Eq.~\eqref{pbh_p} with a top-hat spatial filter, i.e. $W(x)=3(\sin x - x\cos x)/x^3$, gives the velocity field dispersion
\bea
	\sigma_v^2(R,z)
	&=	\frac{3}{10\pi}\left(\frac{1-f_b}{\Omega_m\rho_c}\right)\left[\frac{f(z)H(z)}{1+z}\right]^2\\
	&\times \left(\frac{1+z_{\rm eq}}{1+z}\right)^2\frac{f_{\rm PBH}M_{\rm PBH}}{R}\,.
\eea
Taking a smoothing scale equal to the average PBH comoving distance
\bea
	\bar{ d} = n_{\rm PBH}^{-1/3}
	&=\left[\frac{1}{(1-f_b)\Omega_m\rho_c}\frac{M_{\rm PBH}}{f_{\rm PBH}}\right]^{1/3}\\
	& \simeq 309\,{\rm pc}\left[\frac{1}{f_{\rm PBH}}\frac{M_{\rm PBH}}{\Msun}\right]^{1/3}\,,
\eea
the corresponding 1D velocity dispersion, $\sigma_v^{\rm 1D}=\sigma_v/\sqrt{3}$, then reads
\bea
	\sigma_v^{\rm 1D}(z)
	&=	\frac{(1-f_b)^{2/3}(\Omega_m\rho_c)^{-1/3}}{\sqrt{10\pi}}\\
	&\times	\frac{f(z)H(z)(1+z_{\rm eq})}{(1+z)^2}f_{\rm PBH}^{2/3}M_{\rm PBH}^{1/3}\,.
\eea
For sufficiently large redshifts $f(z)\simeq 1$ and $H(z)\simeq H_0\Omega_m^{1/2}(1+z)^{3/2}$, which leads to
\bea\label{eq:sigma_v_app}
	\sigma_v^{\rm 1D}(z)
	&=	H_0\sqrt{\frac{\Omega_m}{10\pi}}(1-f_b)^{2/3}(\Omega_m\rho_c)^{-1/3}\\
	&\times	\frac{1+z_{\rm eq}}{\sqrt{1+z}}f_{\rm PBH}^{2/3}M_{\rm PBH}^{1/3}\\ 
	&\simeq	 6.0\,{\rm km/s}\frac{f_{\rm PBH}^{2/3}(M_{\rm PBH}/\Msun)^{1/3}}{\sqrt{1+z}}\,.
\eea
For example, $z\simeq 1100$, for $f_{\rm PBH}=1$ and $M_{\rm PBH}=30\,M_\odot$ we obtain $\sigma_v^{\rm 1D}\simeq 0.56$~km/s, which describes the low tail of the velocity distribution obtained from our simulations relatively well.

{We remark that the scaling $(1+z)^{-1/2}$ in Eq.~\eqref{eq:sigma_v} results partly from our assumption that the effective filtering scale $R$ in Eq.~\eqref{eq:sigma_v_def_app} is provided by the mean comoving distance between the PBHs -- one of the principal characteristics of the PBH population. However, for a more precise treatment one could attempt to break the problem of finding $\sigma_{\rm PBH}$ into two parts: (i) linear large-scale motions and (ii) nonlinear motions of PBHs inside clusters, with an appropriate treatment for the scale dividing these two regimes. The latter approach was taken in a recent paper~\cite{Inman:2019wvr} which appeared after this work was completed. Omitting the nonlinear motion gives a conservative estimate for the suppression factor. As a consistency check, we reevaluated the corrections to accretion using the analytic structure formation estimates of Ref.~\cite{Inman:2019wvr}, including the virial motion inside the clusters, and found that our conclusions remain intact.}

%%%%%%%%%%%%%%%%%%%%%%%%%%%%%%%%%%%%%%%%%%%%%%%%%%%%%%%%
\bibliography{PBHaccretion.bib}

\end{document}